\documentclass[12pt]{article}
\usepackage{epsfig,latexsym}
\title{HOLOGRAPHY, CFT AND BLACK HOLE ENTROPY\footnote{Based on
    invited talks at the Platinum Jubilee International
    Conference on Theoretical Physics at ISI,
    Kolkata, December 2007 and the International Conference on Quantum
    Gauge Theories at the SNBNCBS, Kolkata, January 2008.}}
\author{PARTHASARATHI MAJUMDAR \\Theory Group, Saha Institute of
Nuclear Physics \\Kolkata 700064, India.\footnote{E-mail: parthasarathi.majumdar@saha.ac.in}} 
\begin{document}
\maketitle
\begin{abstract}
Aspects of holography or dimensional reduction in gravitational
physics are discussed with reference to black hole thermodynamics.
Degrees of freedom living on Isolated Horizons (as a model for
macroscopic, generic, eternal black hole horizons) are argued to be
topological in nature and counted, using  their relation to two
dimensional conformal field theories. This leads to the
microcanonical entropy of these black holes having the
Bekenstein-Hawking form together with finite, unambigious {\it quantum
spacetime} corrections. Another aspect of holography ensues for
radiant black holes treated as a standard canonical ensemble with
Isolated Horizons as the mean (equilibrium) configuration. This is
shown to yield a universal
criterion for thermal stability of generic radiant black holes,
as a lower bound on the mass of the equilibrium isolated horizon in
terms of its microcanonical entropy. Saturation of the bound occurs
at a phase boundary separating thermally stable and unstable phases
with symptoms of a first order phase transition. 
\end{abstract}

\section{Introduction} 

The laws of black hole mechanics \cite{bard72} relate changes in the area
${\cal A}_{hor}$ of a spatial section of the event horizon (EH) of
stationary black hole spacetimes to variation in parameters like the
change in the ADM mass $M$ and the surface gravity $\kappa_{hor}$ on
the EH,
\begin{eqnarray}
\delta {\cal A}_{hor} ~&\geq&~ 0 \cr
\kappa_{hor} ~&=&~ const \cr
\delta M ~&=&~ \kappa_{hor}~ \delta {\cal A}_{hor} + \cdots ~,
\label{lbhm}
\end{eqnarray}
These laws appear to have a curious analogy with the zeroth,
first and second laws of standard thermodynamics, with the area of the
EH. Yet, black hole spacetimes like Schwarzschild and Kerr emerge as
exact solutions of the {\it vaccum} Einstein equation in complete absence of
matter or energy. There is no conceivable source for
any kind of {\it microstates} usually thought to be as the origin of
thermodynamical behaviour. The originators of these laws,
understandably, did not venture beyond their formulation and derivation
from general relativity.

Bekenstein \cite{bek73} was the first to argue that these laws {\it
must} signify thermodynamic behaviour of spacetimes, beyond mere
analogy. If an object falls through an EH, there is a net reduction in
the entropy of the part of spacetime external to that EH. To be
consistent with the second law of standard thermodynamics, these
spacetimes themselves must carry entropy, whose increase compensates
for the reduction mentioned above, such that the sum of the black hole
and external entropies never decreases. This, essentially, is Bekenstein's
statement of the so-called {\it Generalized} second law of
thermodynamics. In order for this black hole entropy to be consistent
with eq. (\ref{lbhm}), Bekenstein proposed that 
\begin{eqnarray}
S_{bh} ~=~ { {\cal A}_{hor} \over 4 l_P^2 }~ (k_B=1). ~\label{bhal}
\end{eqnarray}
where, $l_P \equiv (G \hbar /c^3)^{1/2} \sim 10^{-33} cm$ is the
Planck length, usually taken to be scale at which quantum
gravitational effects dominate physics. The factor 4 in (\ref{bhal})
emerges from Hawking's formulation \cite{haw75} of black holes in
presence of ambient quantum matter as radiating like a
thermal black body with a temperature given by $\kappa_{hor}$, thus
abundantly substantiating Bekenstein's hypothesis. However, the
natural emergence of the Planck length in (\ref{bhal}) goes to suggest that
thermodynamics associated with classical spacetimes really stems from
microstates arising from the quantum `atoms' of such spacetimes, as
would be found in a theory of quantum gravitation. What remains a
puzzling observation however, is the dependence of black hole entropy
upon an {\it area} rather than on three dimensional volume as in
standard thermodynamics. How and why that happens are the key issues we
wish to address in this article.

The next section deals with a discussion of the Holographic hypothesis
as formulated by 't Hooft and our proposal that it may be a
consequence of the huge invariance group associated with spacetime in
general relativity. This is followed by a description of Weakly
Isolated Horizons (WIH) as an inner boundary of spacetime which serves as 
the protptype of a large class of horizons
including but not only, stationary black hole event horizons. How
holography emerges in this description is also pointed out, especially
in connection with the topological decription of the boundary degrees
of freedom. A very brief description of Loop Quantum Gravity is next
given, with emphasis on the spectrum of certain geometrical
observables. How this leads to a quantum theory of WIH and eventually
to the microcanonical entropy of generic, macroscopic, eternal black
holes (in 4 dimensions) is in the next section. Another version of
holography in spacetime physics is discussed next, within the context
of a standard equilibrium statistical mechanics approach to a canonical
ensemble of radiant (hence non-isolated) black holes. A rather
universal criterion of thermal stability of radiant black holes is
discussed, in which the mass of the equilibrium configuration (chosen
to be a WIH) is
bounded from below by the microcanonical entropy of this
configuration. The point of saturation of this bound corresponds to a
`phase boundary' between a thermally stable and an unstable phase; the
transition has tell-tale signs of a first order phase
transition. Brief comparison with the pioneering work of Hawking and
Page is given. We end with a list of issues yet unresolved.

\section{Holographic Hypothesis}

The Holographic hypothesis was postulated by 't Hooft \cite{thft93} as
one way of understanding how in gravitational physics, information
about the full black hole spacetime is encoded on the EH - a three
dimensional null hypersurface (which is also an `outer trapped
surface'and a boundary of the black hole spacetime). The main idea of 
this hypothesis is best stated in 't Hooft's words \cite{thft93}, 

{\it ... Given any closed surface, we can
represent all that happens inside it by degrees of freedom on this
surface itself. This ... suggests that quantum gravity should be
described by a {\bf topological} quantum field theory in which all
degrees of freedom are projected onto the boundary.}

The questions that immediately arise are
\begin{itemize}
\item Is Holography  in (quantum) general relativity a consequence of
diffeomorphism and local Lorentz invariance of spacetime in presence
of boundaries (like the EH)  ? If it does arise {\it naturally} in
spacetime physics, one need not {\it postulate} it as an additional
hypothesis. 
\item Is there a link to a topological field theory on a boundary of  
spacetime ?
\item Is there any relation to a two dimensional conformal field theory ?
\item How does one compute the entropy of black holes on this basis ?
\item Is there any implication for thermal stability of radiant black holes ?
\end{itemize}

The above list is somewhat biased toward the order of topics in this
article, and hence can be construed as a list of contents. We shall
show, not always rigorously, that the answer to each
of the queries above is most likely in the affirmative, thus obtaining
most of what 't Hooft hypothesized a decade and a half ago. 

We begin with the oft-made observation that local gauge invariance is
{\it not} a statement of symmetry but rather of {\it redundancy}
of some of the
degrees of freedom used to formulate the theory. E.g., in vacuum Maxwell
electrodynamics, the photon field ${\bf A}$ admits the
decomposition 
\begin{eqnarray}
{\bf A} ~&=&~ \underbrace{{\bf A} ~-~ {\bf grad} \int d^4x'G({\bf
    x-x'}) ~{\bf div'~A({x'})}}_{\bf A_T} \nonumber \\
&~+~& {\bf grad} \underbrace{\int d^4x'G({\bf x-x'}) ~{\bf
    div'~A(x')}}_{\bf a_L} ~\label{phot}
\end{eqnarray}
where $\Box~G({\bf x-x'})=\delta^{(4)}({\bf x-x'})$. Under a gauge
transformation ${\bf A} \rightarrow {\bf A}^{\omega} =
{\bf A} + {\bf grad}~\omega$
\begin{eqnarray}
{\bf A_T}~\rightarrow~{\bf A_T}^{\omega} &=&~{\bf A_T} \nonumber \\
{\bf a_L}~\rightarrow~{\bf a_L}^{\omega} ~&=&~{\bf a_L} ~+~ \omega ~, \label{gauge}
\end{eqnarray}
thus clearly revealing the unphysical nature of the longitudinal
degree of freedom ${\bf a_L}$. The 4-vector field ${\bf A_T}$ is
4-divergrence free (transverse in spacetime). Integral curves of this
field are closed spacelike curves. It is trivial to show that the
Maxwell field strength tensor is uniquely determined by ${\bf A_T}$
and is completely independent of ${\bf a_L}$. Yet, the standard
formulation of the Maxwell theory continues to use this redundancy, perhaps
because most feel it is convenient to do so. What is also obvious is
the nonexistence of a gauge invariant tensorial conserved N\"other 
current for vacuum electrodynamics, corresponding to gauge
transformations. If gauge invariance were a symmetry, surely such a
current would be in existence. 

A similar redundancy exists in spacetime physics as portrayed in
general relativity. Spacetime is diffeomorphic to itself under
coordinate diffeomorphisms, thus underlining the redundancy of 
coordinate frames. The clinching evidence for this is the nonexistence
of a covariantly conserved energy momentum tensor for spacetime. All
attempts to construct such a tensor at best yield non-covariant
expressions that can hardly be given a true physical
meaning. There is thus no such thing as a local `gravitational' energy
density in the spacetime of general relativity. 

This state of affairs becomes clearer in a canonical
formulation of vacuum general relativity. In this formalism,
diffeomorphism invariance of the theory leads to a
`bulk' Hamiltonian
\begin{eqnarray}
H_{bulk} ~=~ \int_M \left[ \underbrace{{\cal
      N}}_{lapse} {\cal H} + \underbrace{{\bf N}}_ 
{shift} \cdot {\bf P} \right] ~, \label{hamil}
\end{eqnarray}
where $M$ is a three dimensional spacelike hypersurface (partial Cauchy
surface) on which initial data are specified.  This means that the
Lapse and Shift functions are Lagrange multipliers, 
enforcing the First Class constraints (infinitesimal
spacetime diffeomorphismn generators),
\begin{eqnarray*}
{\cal H}({}^3g,{}^3\Pi) &~ \approx~ & 0 \nonumber \\
{\bf P}({}^3g,{}^3\Pi) &~ \approx~ & 0~. \label{h=0=p}
\end{eqnarray*}
Here, phase space is spanned by the 3-metric ${}^3g$ and 3-momenta
${}^3\Pi$. On the constraint surface then one has $H_{bulk} \approx
0$. There is thus no notion of a `bulk' energy associated with
spacetime in general relativity, signifying that diffeomorphism
invariance is {\it not} a symmetry but a redundancy. In a formulation
of general relativity using orthonormal tetrads as local Lorentz frame
variables, there is in addition, another first class constraint
corresponding to local Lorentz transformations. 

Is there any notion at all of `gravitational energy' in general
relativity? For spacetimes that are asymptotically flat, i.e., those
that in some sense approximate Minkowski spacetime infinitely far away
from the EH, both in space and time, it is possible to define globally
conserved quantities like energy, momentum or angular momentum at the 
boundary of spacetime. These definitions
of global generators depend  crucially on how one describes the
asymptopic structure of spacetime. For spacetimes with additional
`inner' boundaries like the WIH, one can define an energy associated
with the WIH, which is distinct from the energy defined at the
asymptopic boundary. It is obvious in any case that the total
Hamiltonian for a general relativistic system is given on the
constraint surface in phase space by
\begin{eqnarray} 
H~=~H_{bu;l}~+~H_{bdy}~\approx~H_{bdy}~, \label{toth}
\end{eqnarray}
where `boundary' may consist of several disconnected hypersurfaces
embedded in spacetime. Our main interest in what follows will be on
WIHs as inner boundaries of spacetimes. In any event, the very fact
that any notion of gravitational energy or momentum or indeed angular
momentum of spacetime refers to the boundary of spacetime rather than
the bulk is ample evidence of holography at play. 

\section{Weakly Isolated Horizons}
Event horizons of stationary black holes are excessively {\it
global}. This is implied in the following features of such horizons:
\begin{itemize}
\item Event horizons are determined only after the entire spacetime is known. 
\item Stationarity implies that the black hole metric has {\it global}
timelike isometry with the corresponding Killing vector field
generating time translations at spatial infinity. 
\item Event horizons are usualy treated separately from cosmological
horizons like de Sitter horizons. A unifying treatment is desirable.
\item The ADM mass featuring in the First law of black hole mechanics
in eq. (\ref{lbhm}) is defined at spatial infinity and is in no way
associated with the event horizon. 
\end{itemize}
These features make is necessary to seek generalizations which are
characterized locally. 

The particular generalization which we adopt here is known as the {\it
weakly Isolated Horizon} (IH), developed in \cite{ash96}.

\begin{figure}
\begin{center}
\epsfxsize=12cm
\epsfbox{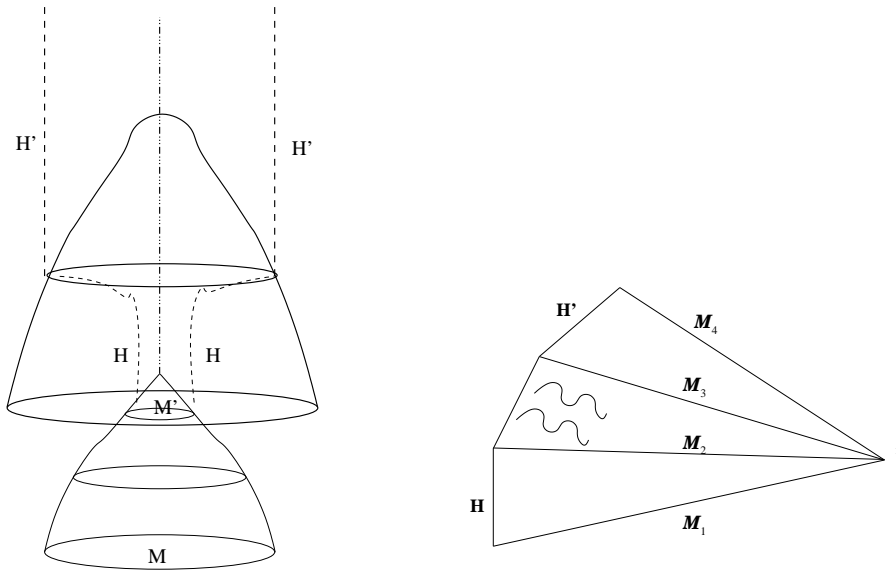}
\end{center}
\caption{Weakly Isolated Horizon}
\label{aba:fig1}
\end{figure}

The properties of an IH can be summarized as follows \cite{ash96}:
\begin{itemize}
\item It is a {\it null inner boundary} of spacetime with topology 
${\bf R} \otimes {\bf S}^2$.
\item The area of the IH $ A_{hor \sim S^2} = const.$ This is what is
inherent in the isolation. It also means that the IH is never crossed,
even though there may exist matter and radiation arbitrarily close to it. 
\item It is a marginal outer trapping surface. 
\item There is no global timelike isometry associated with the IH;
this implies that {\it nonstationary} generalization of stationary black hole horizons.
\item On IH one can define a surface gravity $\kappa_l$, which however
is not defined outside of the IH, since there is no global timelike Killing
vector field allowing such a definition.  
\item The `Zeroth law of IH mechanics' $\kappa_l|_{IH}=const$ can be
demonstrated, although the norm of $\kappa_l$ is not fixed since
there once again there is no timelike Killing vector.
\item On IH, one can define mass a $M_{IH} = M_{ADM} - {\cal E}_{rad}^{\infty} $ such
that $ \delta M_{IH} = \kappa_l \delta A_{hor} + \dots$; this may be
termed as {\it First law of Isolated Horizon Mechanics }
\item Such horizons correspond thermodynamically to a microcanonical
ensemble with fixed $ {\cal A}_{hor} $.
\end{itemize}

Now, because any IH is an inner boundary, the variational principle
cannot be applicable without an appropriate boundary term
\begin{eqnarray}
{\cal S} ~=~ {\cal S}_{EHL}~ +~ {\cal S}_{IH}~  \label{stotal}
\end{eqnarray}
where, the second term is chosen such that its variation cancels the
surface term arising from the variation of the
Einstein-Hilbert-Lorentz action. On the other hand, since an IH is
null, the 3-metric on the IH is {\it degenerate} :
$\sqrt{^3g}=0$. This implies that on the IH one cannot have an action
of the usual type $\int_{{}_{IH}}~ \sqrt{^3g}~ {\cal L}$. Rather, the
quantum field theory describing the IH dof must be a 3 dimensional Topological
Field Theory (TFT). 

The issue that must be addressed now is : which TFT ? This question is of
considerable importance because the microcanonical black hole entropy 
$S_{bh} \equiv \log \dim {\cal H}_{TFT }$, where ${\cal H}_{TFT}$ is
the Hilbert space corresponding to the theory describing the dynamics
of the IH. For this one needs to briefly touch upon the matter of
appropriate canonical variables for general relativity. So far we have
written eqn.s (\ref{hamil}), (\ref{h=0=p}) and (\ref{toth}) 
symbolically in terms of functions of the 3-metric $^3g$ and its conjugate
3-momenta. Through canonical transformations and fixing the gauge
invariance associated with local Lorentz boosts (time gauge), a more
convenient formulation is seen to emerge \cite{imm93} for a {\it real} $SU(2)$
connection ${\cal A}_{SU(2)}$ on the spacelike Cauchy surface $M$ 
chosen to supply Cauchy
data, with its conjugate momenta being the {\it densitized} triads
$E$, the pullbacks of tetrads (local frame fields) to $M$.   

With IH boundary conditions, the solution space of
vacuum Einstein equation admits a closed two-form (symplectic structure)
\begin{eqnarray}
\Omega(\delta_1,\delta_2) & ~=~ & {1 \over 16 \pi l_P^2} tr
\int_{M} ~[\delta_1 {\cal A} \wedge \delta_2 {\bf \Sigma} ~-~ (1
  \leftrightarrow 2) ] \nonumber \\
&~+~& {A_S \over 8\pi \gamma l_P^2} tr \int_{S} ~[\delta_1 {\bf A}
\wedge \delta_2 {\bf A} ~-~ (1 \leftrightarrow 2)] \nonumber \\  
& ~\equiv~ & \Omega_{blk} ~+~ \Omega_{bdy}~, \label{sympl}
\end{eqnarray}
where, $S$ is spatial foliation of IH by $M$ and 
${\bf \Sigma} \sim E \wedge E$. The solution space corresponding to
the $\Omega_{bdy}$ is the one corresponding to the $SU(2)$ Chern Simons 
equation with the pullback of ${\bf \Sigma}$ playing the role of
source on the IH. Further pulled back to the foliation $S$ of the IH, this
equation is the Chern Simons Gauss law
\begin{eqnarray}
\left(\frac{k}{2\pi} {\bf F}_{CS}({\bf A}) ~+~ \Sigma  \right)
|_{{}_{S}} ~=~0~ \label{consis}
\end{eqnarray}
where, $k \equiv [A_S / 8\pi l_P^2] $ with $[]$ signifying `nearest
integer'. Thus, the entire role of bulk spatial degrees of freedom 
characterized by
$\Sigma $ (determined by solving Einstein equation) is as {\it source}
for the Chern Simons degrees of freedom (given by ${\bf F}_{CS}$) 
characterizing boundary (IH) geometry. It is clear that this is a
version of a Holographic picture, albeit in a somewhat subtle way. 
Certain aspects of the Holographic Hypothesis have already been
realized such as primacy of boundary degrees of freedom, a TFT on the
inner boundary (IH) and so on. Note that this realization of the
holography paradigm has been crucially dependent on the diffeomorphism
and local Lorentz invariance of spacetime {\it and} the IH
dynamics. We shall see another aspect of this picture later.

\section{Loop Quantum Gravity : Spin Network Basis in brief}

This is a background-independent, non-perturbative approach to quantum
general relativity\cite{ashlew03}. Quantum three dimensional space is
pictured as a network with edges carrying spin $j=n/2~,~n \in {\cal
Z}$ and vertices consisting of invariant $SU(2)$ tensors
(`intertwiners'). The operators on this space are nonlocal, being
holonomies of the $SU(2)$ connection along edges of the network and
smeared densitized triads. The network is not a rigid network but a
floating one more like a 3 dimensional fishnet. The length of the
edges is not fixed to any scale; nor are the edges required to remain
straight. Arbitrary knottings of edges are allowed. The vertices can
have any valence consistent with conservation of net spin on the edges
meeting at a vertex. Local Lorentz invariance and spatial
diffeomorphism invariance require that the network be a closed one
with no `hanging' edges. Each state in this basis is required to be
annihilated by the local Lorentz and spatial diffeomorphism
generators, and the set of all spinnet states span a {\it kinematical} 
Hilbert space. The physical Hilbert space, consisting of the kernel of
the Hamiltonian constraint, is yet to be worked out in detail. 

The spinnet basis is the eigenbasis for geometrical operators like
length, area and volume in three dimensions. Consider for instance a two
dimensional spacelike surface of classical area ${\cal A}_{cl}$ 
embedded in a spin network; links
of the network will intersect this surface. Assume that the surface is
divided into tiny patches such that each patch is pierced by only a
single link, whose spin is encoded in the puncture on that
patch. The eigenvalues of the {\it area} operator are then given, in
terms of the spins $ j_i, i=1,2, \dots, N$ at these punctures (or 
equivalently, the spin numbers $n_i$) by 
\begin{eqnarray}
a(n_1, \dots, n_N) &=& \frac14 \gamma l_P^2 \sum_{p=1}^N  \sqrt{n_p(n_p+2)} \\
\lim_{N \rightarrow \infty} a(n_1,....n_N) & \leq & {\cal A}_{cl} +
O(l_P^2) ~\label{areaop} 
\end{eqnarray}
The area operator thus has a bounded, discrete spectrum, even though
there is a certain degree of arbitrariness in the scale of
discreteness \cite{ashlew03}. Also, the rate of convergence of the
discrete eigenvalues to the continuum value is approached rather
rapidly \cite{ashlew03}. 
 
\section{ Quantum Isolated Horizon}

Now, the IH embedded in a spatial geometry represented by spin
networks; the spatial section of an IH, assumed spherical here for
simplicity, is a two dimensional surface punctured by spin network
links which transmit their spins to the punctures. A cartoon depicting
this is shown in the following diagram.  

\begin{figure}
\begin{center}
\epsfxsize=12cm
\epsfbox{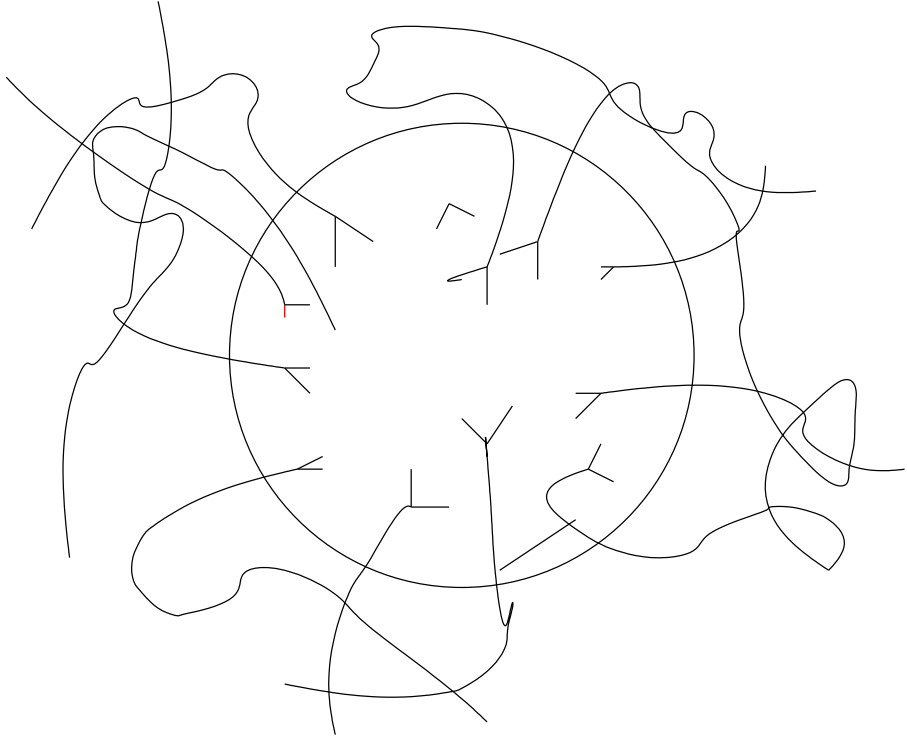}
\end{center}
\caption{Quantum Isolated Horizon}
\label{aba:fig2}
\end{figure}

With the kind of penetration of spin network links discussed above,
the Consistency Condition eq.(\ref{consis}) can be expressed as
a condition in terms of the relevant quantum operators operating on the
kinematical Hilbert space,
\begin{eqnarray}
\left( k{\hat F_{CS}}~+~{\hat E} \times {\hat E} \right)_S
|\Psi\rangle ~=~0 ~, \label{consisq}
\end{eqnarray} 
where 
$|\Psi \rangle \in {\cal H}_{bulk} \otimes {\cal H}_{bdy}$. The object
now is to compute $ \dim {\cal H}_{bdy}$ which is basically the
dimensionality of the CS Hilbert space on an IH with $N$ punctures on
the $S^2$ having spins $\ { j_i / }$ acting as pointlike sources.  

\subsection{Counting of CS states}

Having argued that the boundary (IH) degrees of freedom constitute
those of an $SU(2)$ Chern Simons theory with pointlike spin-valued
sources, we now relate the Hilbert space of the Chern Simons theory to
{\it conformal blocks} of the $SU(2)_k$ WZW model that lives on the
foliation $S$ of the IH, following Witten and others
\cite{wit89}. According to this relationship, $ \dim {\cal H}_{CS+sources} ~=~
\Omega(j_1,\dots,j_N)$ where
$\Omega$ {\bf is the number of {\it conformal blocks of the $SU(2)_k$
WZW model} `living'  on $S^2$ with point sources carrying spins
$j_1,\dots, j_N$}. Using the Verlinde formula, this number can be
computed exactly \cite{km98}. Alternatively, one can solve the Chern
Simons theory and calculate $\Omega(j_1, \dots , j_N)$ directly
\cite{ash97}. Our interest is in macroscopically large black holes
(IHs), i.e., those IHs whose areas $A_{IH} >> l_P^2$. Thus the
deficit angles of punctures made are being required to together 
{\it represent} a smooth $S^2$. Intuitively, this means that one must
maximize the number of punctures for a given fixed classical area, and
this requires the spin on each puncture to be as small a possible,
i.e., spin 1/2. With this choice, $\Omega(1/2,1/2, \dots, 1/2)$ can be
easily computed and its logarithm extracted, giving the microcanonical
entropy \cite{km00}, 
\begin{eqnarray}
S_{micro} = k_B \left[ {{\cal A}_{hor} \over 4 l_P^2} - \frac32 \log \left( {{\cal
A}_{hor} \over 4l_P^2} \right) + O \left( ( {{\cal A}_{hor} \over 4 l_P^2}
)^{-1} \right ) \right]~. \label{smicro}
\end{eqnarray}
where, the correct normalization for the leading oder
Bekenstein-Hawking result emerges as a fit for a real parameter
(Barbero-Immirzi parameter \cite{imm93} invariably multiplying
the Planck scale \cite{ash97}. Observe, however,
that this is the only ambiguity in the entire infinite series each of
whose terms are finite.

Admittedly, the counting presented above is rather crude! One should
actually consider a set of $N$ punctures with a spins $j_i,
i=1,\dots,N$, count $\Omega(j_1, \dots, j_N)$ and sum over all
possible spins and punctures. Setting {\it all} spins equal to 1/2 is
not guaranteed to give the same number, although in view of our
argument regarding deficit angles, it may be a reasonable
approximation for asymptotically large horizon areas. Various
computations of $\Omega$ by this procedure have been reported. The one
that appears to resemble  a truly statistical mechanical computation
is that of \cite{gm06} which is quite consistent with our final result
(\ref{smicro}) in so far as leading logarithmic corrections to the area law are
concerned\footnote{The coefficient of the logarithmic correction
depends upon the particular Chern Simons theory, as explained in
detail in \cite{cm03}}.

\section{Low-tech way : It from Bit}

The idea of {\it It from bit} \cite{whee92, thft93} essentially is to
think of a floating two dimensional lattice covering a two dimensional
sphere $S^2$ which we take to be the horizon. Each plaquette of the
lattice is taken to be of Planck area size, so that the area of the
horizon in Planckian units is given by an integer $p \equiv
A_{hor}/l_P^2$ which is also
the number of plaquettes covering the horizon. Place in each plaquette
a spin 1/2 object so that two states can be associated it, the
$m_s=\pm 1/2$ states. If all states of the horizon with random
orientation of spin 1/2 variables are considered, the number of states
$\Omega(p) = 2^p$. However, as in the previous subsection, we are
interested in states of this system with {\it net spin $j_{tot}=0$} so that 
\begin{eqnarray}
\Omega(p) = \underbrace{\left( \begin{array}{c} p \\ p/2 \end{array}
  \right)}_{m_{tot}=0}  -
\underbrace{\left ( \begin{array}{c} p \\ (p/2 +1) \end{array} \right)
  }_{m_{tot}=\pm 1} ~\label{itbit}
\end{eqnarray}
which yields the same degeneracy and $S_{micro}$ upon using the
Stirling approximation for the factorials in eq. (\ref{itbit}). 

\begin{figure}
\begin{center}
\epsfxsize=7cm
\epsfbox{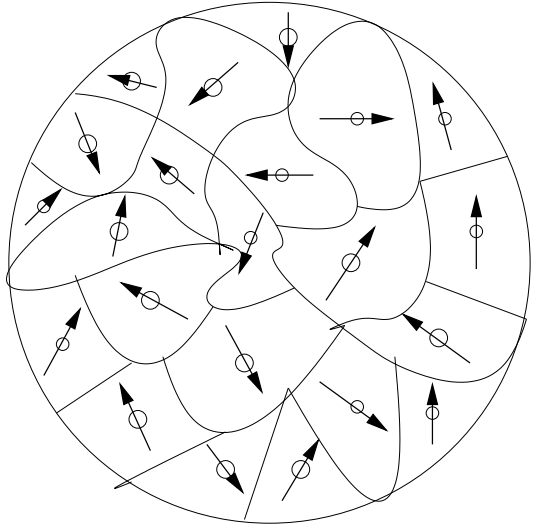}
\end{center}
\caption{It from Bit}
\label{aba:fig3}
\end{figure}

Without having made any use of the Holographic hypothesis, all aspects
of it have thus been seen to be realized in the ab initio computation of the
microcanonical entropy of IHs which we believe serve as good
prototypes of black hole event horizons. The only lacuna in the
computation is that this is actually a computation of the
dimensionality of the {\it kinematical} Hilbert space rather than the
{\it physical} one. There is thus the crucial assumption that the
states we count also belong to the physical Hilbert space. This
remains a grey area because of the difficulties associated with
discerning the semiclassical behaviour of the states spanning the
kernel of the bulk Hamiltonian operator. On the positive side, the
calculation we present yields an asymptotic series of terms of
decreasing powers of the area of the IH, of which only the first,
i.e., the term proportional to ${\cal A}_{IH}$ had been anticipated by
Bekenstein and Hawking. Each of the quantum spacetime correction terms
is robust, unambiguous and finite, requiring no ad hoc regularization
or indeed renormalization of coupling constants. However, the system
under consideration corresponds to a nonradiant (or nonaccreting)
isolated black hole and therefore is unphysical. It is not as yet
clear how the foregoing formalism can be extended or generalized to
realistic situations. In the following we adopt a pragmatic approach
based on equilibrium statistical mechanics of canonical Gibbs
ensembles including Gaussian thermal fluctuation corrections, to reach
some understanding of radiant black holes, based on our approach to
quantum spacetime geometry.

\section{Radiant Black Holes}

Black holes undergoing Hawking radiation (or indeed thermal accretion)
show a runaway behaviour in that as they radiate, their Hawking
temperature increases. This is once again in contradistinction with
standard thermodynamic systems which radiate and cool down as they
approach thermal equilibrium. This instability due to thermal
radiation appears to occur for all asymptotically flat spacetimes
\cite{hawp83} but not necessarily for asymptotically anti-de Sitter
spacetimes. In this part of the article, we analyze the situation from
our standpoint of equilibrium statistical mechanics within a `mean
field' approach incorporating thermal fluctuations. We also choose the
mean field or equilibrium configuration to be an IH whose quantum
behaviour we believe is on firm footing as explained in the previous
sections. Beyond this assumption, our approach is not exactly
semiclassical, in contrast to most of the literature on black hole
thermodynamics, because we do not use any aspect of classical spacetime
geometry, like the form of the metric or even its asymptotic
structure. This allows us to derive a general criterion for thermal
stability of black holes. 

In canonical general relativity, we mentioned that the bulk Hamiltonian
is a sum of first class constraints. The total Hamiltonian for any
spacetime with boundary is thus
\begin{eqnarray}
H~=~H_{bulk}~+~H_{bdy}
\end{eqnarray}
where, $H_{bdy}$ is the Hamiltonian corresponding to all boundaries
including the one at spatial $\infty$ (i.e., the ADM Hamiltonian) such that
\begin{eqnarray}
H_{bulk}~\approx~0
\end{eqnarray}
In any theory of quantum general relativity, one expects 
\begin{eqnarray}
{\hat H}~=~{\hat H}_{bulk}~+~{\hat H}_{bdy}
\end{eqnarray}
such that 
\begin{eqnarray}
{\hat H}_{bulk} |\psi_N\rangle_{blk} = 0 ~\label{hamcon}
\end{eqnarray} 
where $|\psi_N\rangle_{blk}$ are states characterizing bulk space.

Choose as basis eigenstates of the full Hamiltonian 
\begin{eqnarray}
|\Psi\rangle = \sum_{N,\alpha} c_{N, \alpha} |\psi_N \rangle_{blk}
|\chi_{\alpha} \rangle_{bdy}  ~\label{basis}
\end{eqnarray}
With these properties, the canonical partition function
\begin{eqnarray}
Z ~& =&~ Tr ~\exp -\beta ({\hat H}_{blk} + {\hat H}_{bdy}) \nonumber \\
&~=~& \sum_{N,\alpha} |c_{N,\alpha}|^2 \langle \psi_N| \exp-\beta {\hat H}_{blk}
|\psi_N \rangle \langle \chi_{\alpha}| \exp \beta {\hat H}_{bdy}
|\chi_{\alpha} \rangle \nonumber \\
&~=~& \sum_{\alpha} |{\tilde c}_{\alpha}|^2  \langle \chi_{\alpha}| \exp \beta {\hat H}_{bdy}
|\chi_{\alpha} \rangle \nonumber \\ 
&~ = &~ const. ~ \cdot ~Tr_{bdy} \exp -\beta {\hat H}_{bdy} \nonumber \\
& \equiv &~ Z_{bdy}~, \label{bparti}
\end{eqnarray}
where we have used (\ref{hamcon}) for the bulk states
$|\psi_N\rangle$. This has the rather remarkable ramification that {\bf black hole 
thermodynamics is completely determined by the boundary partition function}
This is the holographic picture reappearing in another guise, once
again originating from the Hamiltonian constraint charcterizing
temporal diffeomorphism invariance. Consistent with this picture is
the preeminence of the area of the horizon rather than the volume in
determining the entropy of black hole spacetimes, although this is not
directly follow from our arguments.

\subsection{Saddle Point Approximation}

Assume now that 
\begin{itemize}
\item Boundary in question is a black hole event horizon
\item Eigenvalues of $ {\hat H}_{bdy} : {\cal E}_n = {\cal E}(a_n) $ where,
$a_n = 4\pi \sqrt{3} \gamma l_P^2~ n$
\end{itemize}
These imply that 
\begin{eqnarray}
Z_{bdy} &=& \sum_n~ g({\cal E}(a_n)) \exp \beta {\cal E}(a_n) \\
& \simeq & \int dn~ g({\cal E}(a(n))) \exp -\beta {\cal E}(a(n)) ~for~n>>1 \\
& = & \int d{\cal E}~ \exp ( S_{micro}({\cal E}) -\beta {\cal E} -
\log | {d{\cal E}  \over dn}|)~, \label{bpa}
\end{eqnarray}
where $g({\cal E}(a_n))$ is the degeneracy associated with the area
eigenstate state labelled by $a_n$. 
 
We now make the saddle point approximation, with saddle point chosen
to be ${\cal E}=M_{IH}$,
i.e., equilibrium configuration is chosen to be an {\it isolated
horizon} with a mass $M(A_{hor})$.  

In saddle point approx using standard formulae of equilibrium
statistical  mechanics including Gaussian thermal fluctuations, 
\begin{eqnarray}
S_{canon}~=~S_{micro}(A_{hor})~-~\frac12 \log \Delta \label{scanon}
\end{eqnarray}
where, 
\begin{eqnarray}
\Delta \propto { M''(A_{hor} ) S_{micro}'(A_{hor})
 - M'(A_{hor}) S_{micro}''(A_{hor}) \over M'(A_{hor}) S'^2(A_{hor}) }
\end{eqnarray}

For thermal stability, $S_{canon} \rightarrow~ {\rm{real}}~
\Rightarrow \Delta > 0$; it turns out that this is also a necessary
condition which guarantees the positivity of the heat capacity. This
\begin{eqnarray}
\Delta ~ & > & ~0 \\
\Rightarrow M''(A_{hor}) ~S_{micro}'(A_{hor})~&  > & ~M'(A_{hor})~
S_{micro}''(A_{hor}) ~\label{bound}
\end{eqnarray}
Upon integrating this inequality with respect to area and choosing
appropriate dimensional constants for simplicity, we get
\begin{eqnarray}
M(A_{hor}) ~>~ S_{micro}(A_{hor}) ~,\label{crit}
\end{eqnarray}
a criterion \cite{dmb01}, \cite{cm03} - \cite{cm05} described entirely in terms of
quantities well-understood within the IH-LQG framework. 

Since this criterion has been obtained without any reference to
classical spacetime geometry, it is perhaps worthwhile to check that
indeed it holds for known semiclassical situations. 

\subsection{Schwarzschild Black Hole}

In this case, the area and mass are related by the well-known relation 
\begin{eqnarray}
 M(A_{hor}) ~\sim~ A_{hor}^{{}^{1/2}} ~\label{schw} 
\end{eqnarray}
It is obvious that, since the $S_{micro} \sim A_{hor}$ in this
case, for large $A_{hor}$ the bound is not satisfied,
and we expect a thermally unstable situation, as indeed it is. This is
consistent with the heat capacity $C<0$. Unfortunately, this sort of
instability is endemic to all asymptotically flat black hole
spacetimes, as has been discussed in detail in \cite{hawp83}. 

\subsection{AdS Schwarzschild Black Hole}

Asymptotically anti-de Sitter spacetimes have a {\it timelike}
infinity requiring specification of {\it incoming} data. The incoming
radiation has to precisely cancel the outgoing one in order to
completely specify Cauchy data, for a range of black
hole parameters (mass $M$ and cosmological constant $\Lambda$),
thereby guaranteeing a stable thermal equilibrium. This range is given
by $l \equiv (-\Lambda)^{-1/2} << (A_{hor}/4\pi)^{1/2}$. How do we see that
in the criterion derived above ? For this we need only use the
mass-area relation for AdS Schwarzschild black holes
\begin{eqnarray}
M(A_{hor}) ~=~\frac12 \left( {A_{hor} \over 4\pi
}\right)^{1/2}~\left(1 + {A_{hor} \over
4\pi l^2} \right)~. \label{adschw} 
\end{eqnarray}
So long as the area is within the range specified above, it is obvious
that $M \sim A_{hor}^{3/2} > S_{micro}(A_{hor})$ which means that the
inequality (\ref{crit}) is satisfied. As one approaches the endpoints
of the range, i.e., when $l$ approaches $(A_{hor}/4\pi)^{1/2}$, the
system looks more and more like an asymptotically flat Schwarzschild
spacetime, and thermal instability begins to set in. 

It is obvious however that enroute to instability, the inequality
(\ref{crit}) must {\it saturate} for certain values of the
parameters. This can be shown\cite{pm07} to correspond to the heat capacity
blowing up from the positive side. When the inequality reverses,
the heat capacity is definitely negative, exhibiting a sort of
behaviour reminiscent of first order phase transitions in statistical
mechanics.   

As emphasized earlier, no classical metrics
have been used anywhere in this analysis, and in this sense the
treatment here can be thought of as a generalization of the pioneering 
semiclassical treatment
of Hawking and Page \cite{hawp83}. Observe also that the criterion for 
thermal stability involves the
domination of the black hole mass over the microcanonical entropy
which has to do with the disorder associated with the degrees of
freedom of the quantum isolated horizon. Thus the transition from a
thermally stable to an unstable phase, although not a typical phase
transition in statistical mechanics, is still not without similar
symptoms.

\section{Questions Yet to be Resolved}

In this section we list a list of pending issues which await
satisfactory resolution:
\begin{itemize}
\item The origin of the assumed mass-area relation remains somewhat
obscure, although an approach may be to derive it from an analogue of
the relation in LQG between the area operator and the bulk Hamiltonian
\cite{thie99}. 
\item An important issue is the possibility of Hawking radiation from
an IH. It is to be seen what precisely among the characterizations of
isolation can best be discarded to enable this.
\item Can one go beyond effective description of black hole
as IH (inner boundary of sptm) and consider realistic dynamical collapse ?
\item Is the thermal nature of the Hawking radiation spectrum an artifact
of  the semiclassical approximation ?
\item Does the lowest area quantum $\sim l_{{}_P}^2$ have implications 
for the {\it information loss problem} ?
\end{itemize}


\section{References}

\begin{thebibliography}{99}
\bibitem{bard72} J. Bardeen, B. Carter and S. W. Hawking,
Comm. Math. Phys. {\textbf31}, 161 (1972).
\bibitem{bek73} J. Bekenstein, Phys. Rev. \textbf{D7}, 2333 (1973).
\bibitem{haw75} S. W. Hawking, Comm. Math. Phys. \textbf{43}, 199 (1975).
\bibitem{thft93} G. 't Hooft, ``Dimensional Reduction and
Quantum Gravity'' in \emph{Salamfstschrift} edited by A. Alo and
S. Randjbar-Daemi, ICTP, Trieste, 1993; see also R. Bousso,
Rev. Mod. Phys. {\bf 74} 825-874, 2002 for a more recent review. 
\bibitem{ash97} A. Ashtekar, J. Baez, A. Corichi and K. Krasnov,
Adv. Theor. Math. Phys. \textbf{4}, 1 (2000) and references therein.
\bibitem{km98} R. Kaul and P. Majumdar, Phys. Lett. \textbf
{B439}, 267 (1998). 
\bibitem{km00} R. Kaul and P. Majumdar, Phys. Rev. Lett. \textbf{84},
5255 (2000). 
\bibitem{dmb01} S. Das, P. Majumdar and R. K. Bhaduri,
Class. Quant. Grav. \textbf{19}, 2355 (2001).
\bibitem{cm03} A. Chatterjee and P. Majumdar, Arxiv:hep-th/0303030.
\bibitem{cm04} A. Chatterjee and P. Majumdar, Phys. Rev. Lett. 
\textbf{92}, 141301 (2004).
\bibitem{cm05} A. Chatterjee and P. Majumdar, Phys. Rev. \textbf{D
72}, 044005 (2005).
\bibitem{pm07} P. Majumdar, Class. Quant. Grav. \textbf{24}, 1747 (2007).
\bibitem{hawp83} S. W. Hawking and D. N. Page, Comm. Math. Phys. 
\textbf{87} 577 (1983).
\bibitem{ash96} A. Ashtekar, C. Beetle and S. Fairhurst, Class. 
Quant. Grav. \textbf{17}, 253 (1996) and references therein.
\bibitem{imm93} G. Immirzi, ArXiv:gr-qc 97010052 and references
therein.
\bibitem{ashlew03} A. Ashtekar and J. Lewandowski, Class. Quant. Grav. 
\textbf{\bf 21} R53 (2003).
\bibitem{wit89} E. Witten, Commu. Math. Phys. \textbf{151}, 321 (1989).
\bibitem{gm06} A. Ghosh and P. Mitra, Phys. Rev. \textbf{D 74}, 064026
(2006).
\bibitem{whee92}  J. A. Wheeler, ``It from Bit'' in \emph{Sakharov
Memorial Lectures} Vol. II, Nova Publishing, Moscow, 1992.
\bibitem{thie99} T. Thiemann, Phys. Lett. \textbf{B380}, 257 (1999).
\end{thebibliography}
\end{document}